\documentclass[twocolumn,floats,aps]{revtex4}
\usepackage{times,amsmath,psfrag,latexsym,pstricks,graphics}

\begin{document}
\title{Energy Loss from a Moving Vortex in Superfluid Helium}
\author{R.J. Zieve, C.M. Frei, and D.L. Wolfson}
\affiliation{Physics Department, University of California at Davis}
\begin{abstract}
We present measurements on both energy loss and pinning for a vortex
terminating on the curved surface of a cylindrical container.  We vary
surface roughness, cell diameter, fluid velocity, and temperature.
Although energy loss and pinning both arise from interactions between the
vortex and the surface, their dependences on the experimental parameters
differ, suggesting that different mechanisms govern the two effects.
We propose that the energy loss stems from reconnections with a mesh
of microscopic vortices that covers the cell wall, while pinning is
dominated by other influences such as the local fluid velocity.
\end{abstract}
\maketitle

\section{Introduction}

The entropy of a superfluid flow is entirely contained in its
excitations, the most striking of which are the quantized vortex lines. A
question in many situations is how energy transfer and dissipation occur
within a superfluid.  Energy can shift from one length scale to another,
or between the fluid and macroscopic objects such as the container or an
object moving through the fluid.  Energy can also be dissipated as
phonons or other excitations within the fluid.

Much of the recent work on energy transfer has centered on superfluid
turbulence, where experiments indicate that different mechanisms act in
various temperature regimes. At high temperatures, the superfluid coexists
with a normal fluid. Turbulence in the latter has the standard behavior of
classical turbulence, and through a coupling between the two components
the superfluid takes on the classical behavior as well \cite{Tabeling,
Smith, Walmsley07}. Yet the same power-law behavior of vortex line
density as a function of time, albeit with an altered prefactor, also
applies at lower temperature, where only a negligible amount of normal
fluid remains and its coupling to the superfluid is drastically reduced
\cite{McClintock}.  Only when the method of injecting energy into the
flow changes does the functional form itself change \cite{Walmsley08,
Tsubota}.  On the other hand, while the velocity field in classical
turbulence follows a Gaussian distribution with only minor deviations
appearing three standard deviations from the peak \cite{Noullez}, recent
measurements of superfluid turbulence find non-Gaussian behavior with a
$1/v^3$ form beginning about one standard deviation from the
peak \cite{Paoletti}.  The degree to which superfluid turbulence mimics
classical behavior remains an open question.

Vortex reconnections play a major role in superfluid turbulence.  When two
vortices closely approach each other, they form cusps that are drawn further
together at the tips.  Eventually the connectivity at the tips of the cusps
changes; effectively each cusp is divided in half and one side of a cusp
connects to half of what was originally the other cusp.  Similar behavior
occurs when a vortex closely approaches a wall of a container.  A cusp again
appears which splits in two, and each portion terminates on the wall. The
cusps created in the reconnection process induce Kelvin waves along the
vortices.  The Kelvin waves can transfer energy to smaller length scales or
radiate energy as sound \cite{Vinen01, Charalambousreview}, and in some
cases the increased motion may lead to further reconnections \cite{Ingrid}. 
Reconnections in superfluid helium have only recently been visualized
\cite{Bewley} and are also difficult to simulate since they involve small
length scales and large velocities \cite{Koplik, Tebbs}.  Hence different
types of observations of reconnections are useful for better understanding
the phenomenon.  Here we describe how reconnections may affect the energy
loss from a single moving vortex.

Our apparatus consists of a straight wire, stretched parallel to the axis
of a cylindrical tube \cite{smooth}. The wire can serve as the core of a
superfluid vortex.  Alternatively, a vortex can use the wire as its core
from one end of the cylinder to somewhere in the middle, then leave the
wire and continue to the side wall of the cylinder as a free vortex.  The
wire's vibration frequencies, which we monitor with an electromagnetic
technique, are sensitive to the exact spot where the vortex detaches from
the wire. We can observe various aspects of the motion of the free
portion of the vortex through effects on the detachment point.  For
example, in  this geometry the free vortex precesses around the wire,
driven by the flow field of the trapped circulation.  The displacement of
the wire from the axis of the cylinder forces the length of the free
vortex to change during the precession, which leads to oscillations
of the detachment point that conserve the total energy stored in the
vortex \cite{heli}.

From our previous studies of vortex precession, the
energy loss rate is many orders of magnitude larger than expected
from bulk mutual friction as the vortex moves through the superfluid
\cite{RR}.  A natural assumption is that the dissipation instead
originates from the contact between the vortex and the wall of the
container. An additional observation confirms the presence of a
significant vortex-wall interaction: occasionally a precessing vortex
pins on the wall, producing a characteristic signature that includes
oscillations with higher frequency and smaller amplitude than those
associated with precession, accompanied by an abrupt cessation of the
energy loss \cite{helpin}. Computer simulations, using the assumption that
the wall end of the vortex suddenly stops moving, reproduce all these
features. Very plausibly, forces strong enough to interrupt the vortex
motion completely could also in a less extreme situation induce energy
loss. Here we present further experimental results on the interaction of
a vortex line with the wall.  Our new measurements suggest that different
mechanisms produce the pinning and the dissipation, with reconnections
responsible for the latter.

\section{Experimental Setup}

\begin{figure}[tbh]
\scalebox{.5}{\includegraphics{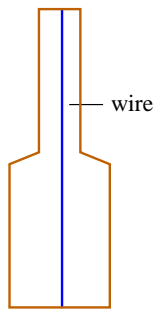}}
\caption{Schematic of typical cell, with diameter changing near the
center.}
\label{f:cell}
\end{figure}

For the present measurements we designed new cells that allow us to explore
the influence of smoothness more thoroughly and also to probe the effect of
cell radius on the vortex-wall interaction.  Our previous work showed
\cite{smooth} that the wire mounting can significantly affect the precession
behavior.  For example, the dissipation is especially high when the wire is
far off-center.  We also know that for a given wire the energy loss depends
on the amplitude of the wire's motion, so other details of the wire's
vibration are likely to play a role as well.  In particular the vibration
frequency varies significantly among wires, from 124 Hz to 791 Hz for the
wires discussed here.   The dissipation rates that we observe for various
wire suggest that other mounting factors which we have not identified also
come into play.  To enable direct comparison between wall treatments, we
make cells that change inner diameter halfway along their length, as shown
in Figure \ref{f:cell}.  

Using a single wire and comparing behavior between the two halves of the
same cell eliminates any dependence on the wire's location, its normal
modes,  the magnetic field needed to excite the vibration properly, and the
excitation pulse's shape and amplitude.  These latter factors determine the
initial motion of the wire, which subsequently moves under the influence of
the trapped circulation. As noted above, the wire's motion itself affects
the energy loss from the moving vortex; hence keeping the wire velocity
constant through a set of measurements is important for meaningful
interpretation.

We identify which half of the cell the vortex is in from the precession
period.  The flow field around the wire drives the precession, and a cell of
larger radius includes fluid that is more distant from the wire and hence
not moving as fast.  This reduces the average speed of the flow, so the free
vortex moves more slowly around the cell. In fact the precession period
depends on the square of the local diameter, which produces a significant
change when the vortex moves from one portion of the cell to the other.

With these cells we can observe directly how cell diameter influences
dissipation.  We can also investigate the effect of wall roughness,
since the diameter change makes polishing only one half of the
cell straightforward.  We do the polishing mechanically, inserting a
Q-tip with diamond powder by hand while rotating the cylinder in a drill
press chuck. We use down to 3 micron diamond powder, resulting in surface
roughness of less than 50 nm.  We test roughness by cutting open polished
cells longitudinally and using a surface roughness comparator. The
measured 50 nm finish agrees with typical polishing results on other
materials, where ultimate surface roughness is typically one or two
orders of magnitude smaller than the abrasive grain size \cite{Tam,
Heydemann, Xu}.

\section{Kelvin Waves}

In our previous work \cite{helpin}, we found several factors that influence
the energy loss from the precessing vortex.  Dissipation increases with
increasing temperature, longer time between measurements, and lower
excitation of the wire during measurements.  All of these effects seem to
have a common source, an influence of the wire's vibration on the
dissipation: the faster the wire moves, the slower the rate of energy loss
from the vortex.  We hypothesized that the wire influences the
dissipation by inducing Kelvin oscillations along the portion of the vortex
stretching between the wire and the cell wall.  Here we confirm that the
wire does indeed excite Kelvin waves along the vortex.

In our new cells, vortices regularly pin at the lip in
the middle of each cell where the inner diameter changes.  Once pinned,
dislodging a vortex is extremely difficult.  Unlike other wall pins,
where vortices often come free unassisted or after vibrating the wire
with larger amplitude than usual, the lip pins rarely work themselves
free. Depinning requires comparable perturbation, either mechanical or
thermal, to depinning from the end of the container for a vortex that runs
along the wire for the entire length of the cell.  The stability of the
pin lets us monitor how varying the wire's vibration amplitude affects
the pinned vortex.

\begin{figure}[t]
\scalebox{.45}{\includegraphics{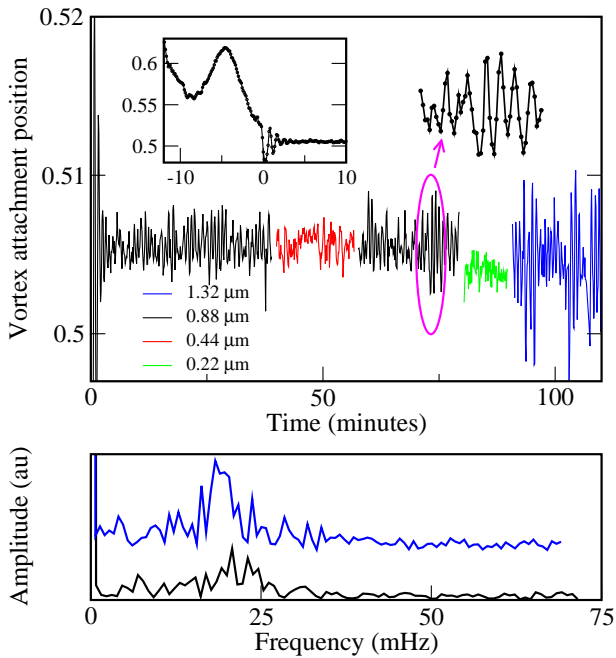}}
\caption{(a) Attachment point position of pinned vortex, for
several vibration amplitudes.  The legend lists the maximum
displacement at the center of the wire, in microns, for each data segment. 
Left inset: precession of vortex before it pins.  Right inset: expanded view
of the fluctuations in the attachment point location, showing a
characteristic frequency. (b) Magnitude of the Fourier transform of the
attachment position, for 0.88 $\mu$m wire excitation (black) and 1.32 $\mu$m wire excitation (blue).  The
curves are shifted from each other vertically for clarity.}  
\label{f:oscamp} 
\end{figure} 
 
Figure \ref{f:oscamp}a tracks a pinned vortex as we vary the excitation
amplitude of the wire.  The legend give maximum displacements at the
center of the wire, in microns.  The variation in the measured attachment
position {\em increases} for larger vibration amplitudes.  This is
exactly opposite from the expected behavior of our measurement error,
since a higher wire excitation typically improves the signal-to-noise as
we identify the wire's vibration frequencies.   One exception is when
the wire is so distorted from its ideal shape that higher-frequency
modes become significant as well as the low-frequency oscillations we
analyze; another is if the oscillations from one measurement do not
die away completely before the wire is excited again. By examining
the traces resulting from individual excitations of the wire, we have
verified that neither of these potential problems is an issue here.
Indeed, as expected, the error in the curve fitting we do to extract
the vibration frequencies goes down as the wire excitation increases.

A more careful examination of the fluctuations about the pin level shows
that they are not random noise, but have a characteristic oscillation period
of about 45 seconds.  The expanded picture of the right inset clearly shows
several measured points per period.  Figure \ref{f:oscamp}b presents Fourier
transforms of the attachment position for the two largest wire excitations,
which produce maximum displacements of 0.88 and 1.32 $\mu$m.  Both have
peaks near 22 mHz, the frequency corresponding to 45 seconds, with a
stronger peak for the larger excitation.  Thus the increased variation about
the pin level is actual motion of the vortex line rather than noise, and the
most natural possibility is the lowest Kelvin mode. Since the vortex is
pinned at the cell wall but free to move along the wire, we calculate the
period of the Kelvin oscillation with quarter-wavelength equal to the cell
radius.  For a wave on an infinite straight vortex, the period is
approximately  $$T = \frac{2\lambda^2}{\kappa \ln(\lambda/2\pi a)},$$ where
$\kappa=9.97\times 10^{-4}$ is the circulation quantum, $a=1.3\times
10^{-8}$ cm is the core radius of a free vortex, and $\lambda$ is the
wavelength of the Kelvin oscillation. These data are from wire D$^\prime$,
with small-side radius 0.16 cm, which yields an ideal Kelvin wave period of
52 s, fairly close to the observed value of 45 s.  The larger-amplitude
oscillations immediately after the pin starts do have period 52 s.  The
period may be reduced to 45 s for subsequent Kelvin waves if the vortex pin
site shifts slightly into the transition region between the two cell
diameters, where the radius is larger.  The amplitude increase with wire
oscillation amplitude and the agreement with the expected Kelvin wave period
show that the wire itself does excite Kelvin waves along the free portion of
the vortex.

\section{Dissipation}

\begin{figure}[tbh]
\scalebox{.49}{\includegraphics{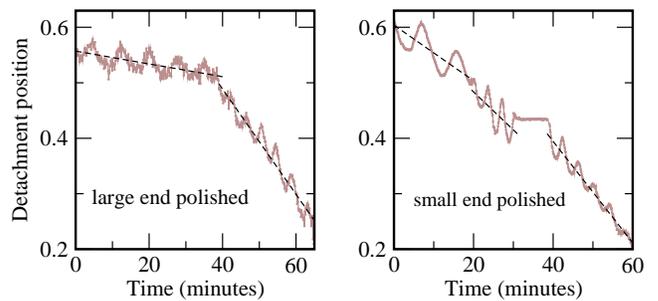}}
\caption{Vortex precession, moving from the large-diameter end
to the small-diameter end, in two cells.  The dashed black lines are
fits indicating the slope of the different precession segments.} 
\label{f:twodiam}
\end{figure}

We next present data on the energy loss during precession in cells with
one half polished. In Figure \ref{f:twodiam}a, the two halves of the cell
have radius 1.6 mm and 2.9 mm, with the larger end polished. 
Precession begins in the wider half of the cell, with a period of 485
seconds.  At the halfway point, the detached portion of the vortex enters
the narrower part of the cell and the precession period shifts abruptly
to 252 seconds, reflecting the faster average velocity field in this
region. The downward slope of the precession trace indicates the steady
decrease in length of the trapped vorticity, which corresponds to a
significant energy loss.  The trapped vortex stores energy per length
$\frac{\rho\kappa^2}{4\pi}\ln\frac{R}{r_w}$, where $\rho\approx 0.145$
g/cm$^3$ is the superfluid density, $\kappa=9.97\times 10^{-4}$ cm$^2$/s
is the quantum of circulation, $R$ is the cell radius, and $r_w\approx
8.7 \mu$m is the radius of the wire.  For this cell, the energy per
length of a trapped vortex is $6.66\times 10^{-8}$ erg/cm at the larger
radius and $5.98\times 10^{-8}$ erg/cm on the smaller side.  Initially
the slope is 0.14\% per minute on the large-diameter side. For our 5 cm
long wire this converts to 7.8$\times 10^{-12}$ erg/s. In the thin half
of the cell, the much steeper slope corresponds to energy loss of
47$\times 10^{-12}$ erg/s.

Previous experimental \cite{smooth} and computational \cite{Schwarzheli}
work has found that changes in energy loss and precession period are
related, with an increase in dissipation corresponding to a longer period. 
However, that relationship holds for precession within a cell of fixed
diameter, not for the change in precession period with cell diameter that we
observe here.  The present relationship between precession period and
dissipation actually goes in the opposite direction: the faster energy loss
occurs when the period is longer.  Thus the previously known correspondence
between dissipation and precession period does not explain our present
observations. 

Figure \ref{f:twodiam}b is less dramatic.  This cell also has diameters
1.6 mm and 2.9 mm, but with the narrow end polished.  Again the vortex
begins in the wider half, and again the change in precession period
near the attachment position of 0.5 indicates its entry into the
smaller part of the cell.  Here too there is a change in the
dissipation rate, although it is a much smaller shift from 28$\times
10^{-12}$ erg/s to 37$\times 10^{-12}$ erg/s.  After about 2.5 periods
on the narrow end, the vortex pins to the cell wall, depinning by itself
after a few minutes.  The dissipation rate increases to
46$\times 10^{-12}$ erg/s after the pin. 

\begin{figure}[tbh]
\scalebox{.48}{\includegraphics{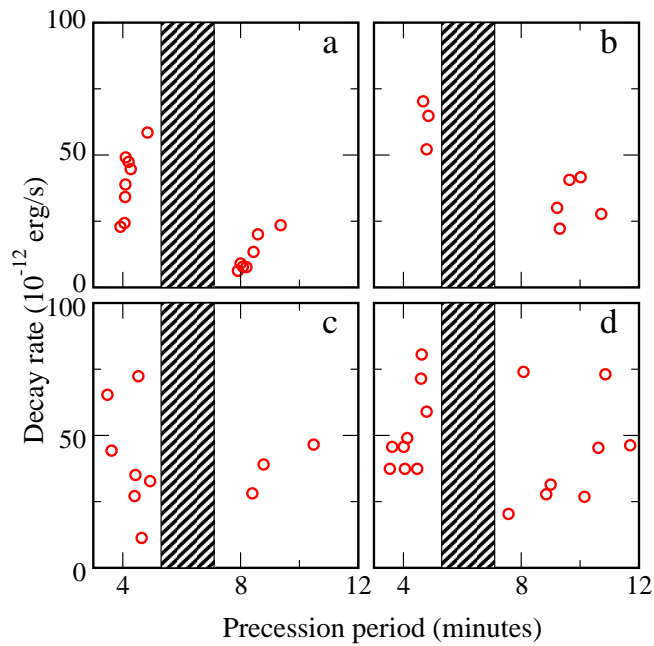}}
\caption{Energy loss rates as function of precession period. 
Shaded regions show the gap in the precession periods that divides the
small-end and large-end motion.  Original (a) and remade (b) cell with only
large end polished; original (c) and remade (d) cell with only small end
polished.}
\label{f:2diamrates}
\end{figure}

The precession periods and energy loss rates of Figure \ref{f:twodiam} are
typical for these two cells.  Figure \ref{f:2diamrates} is a compendium of
results from all the observed precession events. Figure \ref{f:2diamrates}a
shows data from the cell of Figure \ref{f:twodiam}a with the wide end
polished, while Figure \ref{f:2diamrates}c has the narrow end polished. The
decay rates are plotted as a function of precession period.  They segregate
cleanly into narrow-end precession, with period less than 5 minutes, and
wide-end precession, with period more than 7 minutes.  No precession periods
lie in the intermediate region.  For both cells, the dissipation is less in
the wide portion.  In addition, for a given diameter the dissipation is less
on average in the cell with that side polished.  For the cell of Figure
\ref{f:twodiam}a the influence of diameter and smoothness act in the same
direction, and the smooth wide side has dramatically less dissipation than
the rough narrow side.  On the other hand, the cell of Figure
\ref{f:twodiam}b shows much less distinction between its rough wide side and
smooth narrow side. 

To test that other features of the wire mounting, such as the wire's
cross-section, its exact location within the cell, or the angle at
which the wire enters the stycast cap, are not dominating our
observations, we dismantled and remade each of the two cells from the
same brass cylinders.  All the stycast pieces, as well as the wires,
were remade.  We did not do any further polishing to the brass cell
bodies before reassembling the cells. The precession and decay rate
results for the remade cells are shown in Figure \ref{f:2diamrates}b
and \ref{f:2diamrates}d, with the latter including the data from Figure
\ref{f:twodiam}b.  They show the same trends as for the original cells.
The effect of the polishing is less pronounced in the remade cells,
possibly because of accumulations on the surfaces over time.

Two additional cells with a diameter change in the middle, G and H in
Table \ref{t:wires}, also exhibit more dissipation on the narrow side.
These cells have both ends polished, and the contrast between the two
ends is less strong than in cell D but stronger than in cell C.  This
supports the interpretation that both cell diameter and wall smoothness
affect the energy loss.

One known effect of cell diameter is that fluid velocity at the surface
decreases with increasing diameter, a result of the $1/r$ velocity
dependence of the circulation trapped around the wire.  While a lower
velocity could plausibly lead to reduced dissipation, one of our
measurements suggests that this is not the case.  We occasionally see vortex
precession in the regime where one quantum of circulation is completely
trapped along the wire and a second quantum partly covers the wire.  The
result is precession at a circulation value between $N=1$ and $N=2$, with
the vortex driven by a flow field roughly three times as fast as for
circulation between $N=1$ and $N=0$.  The driving field comes from the
trapped vorticity, which in the latter case can be approximated as a
half-infinite vortex.  For circulation between $N=1$ and $N=2$, the
equivalent approximation uses three half-infinite vortices: two running in
one direction from where the vortex detaches and the third running in the
other direction.  This factor of three has experimental confirmation in the
much shorter precession periods when $N>1$. However, the energy loss per
time during precession for $N>1$ is comparable to that for $N<1$.  Hence the
fluid velocity must not be a key factor in the energy loss.

\begin{table*}
\caption{Summary of wires measured.}
\label{t:wires}
\begin{tabular}{l|l|c||c|c|c|c||c|c||c|c|l}
& & &\multicolumn{4}{c||}{number of pin events}&
\multicolumn{2}{c||}{precession time (min)}&\multicolumn{3}{c}{minutes of
precession per pin}\\
cell & radius (mm) & polished & wide end & narrow end & middle & unknown
& wide end & narrow end & wide end & narrow end & temperature\\
\hline
A & 1.6 & yes & -- & 2 & --&--&-- & 219 & -- &110&all below 350 mK\\
B & 1.6 & yes & -- & 0 & -- & --&-- & 193 & -- &$>$193 &300-400 mK\\
C & 1.6, 2.9 & 1.6 & 2& 0&5&0&465 & 255 &232 &$>$255& 350-500 mK\\
C$^\prime$ & 1.6, 2.9 & 1.6 & 3 & 7 & 2&0&225 & 143 &75 &20&400 mK\\
D & 1.6, 2.9 & 2.9 & 0& 0&1&1&430 & 221 &$>$430 &221 &375-400 mK\\
D$^\prime$&1.6, 2.9&2.9&3&1&7&0&262 & 168 &87 &168 &400 mK\\
E & 1.6, 2.9 & no & 2 & 0 & 0 & 18&38 & 35 & 19  &$>$34&mostly 400mK\\
F & 1.8, 1.9 & both & 0 & 0&1&0&-- & 509 & -- &$>$509 &400 mK \\
G &1.8, 3.2&both&9&3&26&0&1366&620&152 &207 &350 and 400 mK\\
H &1.8, 3.0&both&0&1&13&3&482&414&$>$482&414&mostly 400 mK\\
\end{tabular}
\end{table*}

\section{Pinning}

\begin{figure}[tbh]
\scalebox{.4}{\includegraphics{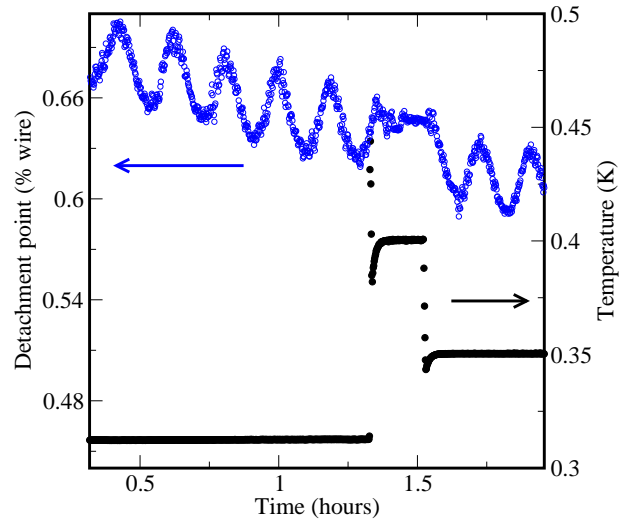}}
\caption{Precession oscillations and pin (blue,
left axis) and cryostat temperature (black, right axis).  The pin
occurs just after the temperature is raised from 312 mK to 400 mK, and
the vortex depins shortly after the temperature is lowered to 350 mK.  Data are from wire G.}
\label{f:lowTpin}
\end{figure}

We next turn to how smoothness and diameter affect pinning.  Table
\ref{t:wires} compiles pinning statistics for several wires.  Cells A
and B have only a single radius along the entire length; the remainder
have a change in diameter as described above.  Since the radius of
cells A and B is typical of the smaller end of the other cells, we
treat all results from those cells as being in the smaller-diameter
region.  Cells C$^\prime$ and D$^\prime$ are the remakes of cells C and D,
and the decays pictured in Figure \ref{f:twodiam} come from wires
C and D$^\prime$. We examine the vortex motion during decays that include a
pinning event. As long as there is a sufficiently clear stretch of
precession, we can determine from its period which half of the cell
contains the pin site. For example, Figure \ref{f:lowTpin} shows a pin
surrounded by precession oscillations of period about 11 minutes.  The
long period indicates that the vortex is on the large-diameter portion
of the cell. We assume that any pin close to the center of the cell is at
the lip. In some cases the precession is too noisy or too brief to
identify where the vortex is.  This is mainly an issue with cell E,
which was completely unpolished and had very frequent pins. 

The pinning properties of the cells described here do not exhibit a
correspondence between increased pinning and higher energy loss.
More pinning occurs in the larger-diameter portions of cells, despite  
the fact that the energy loss is lower in these regions.  Note that
the absolute number of pins is not the correct quantity to compare across
cells; the amount of time used to acquire data in each case is relevant,
as is the duration of smooth precession atop which pin events can be
identified.  The final sections of Table \ref{t:wires} list the total precession
time for each half of each cell and normalize the pinning events to the
precession time. We also list the precession temperatures
in the final column, since the likelihood of pinning does increase with
temperature.

We find large variation in the pinning characteristics from one cell to
another with nominally similar properties.  As far as roughness, which
increases the energy loss rate, there is generally more pinning in rougher
cells.  In an extreme example, the vortex pinned so frequently in the
unpolished cell E that no precession ever continued for more than one and
one-half circuits of the cell.  The only two cells showing more pinning on
their polished ends than on their unpolished ends were the two remade cells,
C$^\prime$ and D$^\prime$.  We speculate that accumulations of dust on the
walls between the original assembly of these cells and the reassembly might
have a larger effect on the polished surfaces. It is possible that dust is a
larger factor for these two cells, since before the second assembly we
cleaned them only with liquid, in an attempt to avoid adding any additional
scratches to the surfaces.

Another indication of the propensity to pin comes from the temperature required
for reliably pinning a vortex.  In previous work \cite{helpin}, we found
that a vortex quickly pins at temperatures above 1 K, coming free if and
when the temperature again drops below 500 mK.  Figure \ref{f:lowTpin} shows a
similar effect on wire G, albeit at lower temperature.  
An increase in temperature from 312 mK to 400 mK triggers the pinning, and the
vortex frees itself once the temperature is lowered to 350 mK.
Because the signal-to-noise in our measurements is much higher in
this temperature range than it is above 1 K, we can see clearly that
the precession oscillations cease at 400 mK.   The role of elevated
temperature in instigating pinning seems qualitatively the same, but
the pinning regime now begins at lower temperatures.  We conclude that
pinning is easier in the new cell.  Indeed, we reduced the temperature
from 400 mK to 350 mK for many of our precession measurements, since
at 400 mK there was so much pinning that we rarely observed significant
stretches of precession.

These results expose a difficulty with the idea that Kelvin waves control
dissipation through their interaction with the wall roughness. If energy
loss is caused by near-pins that the vortex breaks free from, then energy
loss and pinning should track each other; if low dissipation indicates
few near-pinning events, then one would expect correspondingly few
actual vortex pins. Previous observations supported the relationship
\cite{helpin}.  For example, increasing temperature leads both to higher
dissipation and to a much higher probability of pinning. However, the correspondence does not extend throughut the present work.  We summarize
our results on both pinning and energy loss in Table \ref{t:exptsum}.
Notably a larger cell diameter reduces dissipation but
increases pinning, while polishing the cell walls reduces dissipation but
has a less straightforward effect on pinning. In both cases the change
in dissipation is much more reproducible than the effect on pinning.
These observations suggest that the two phenomena, while both involving
interaction between the precessing vortex and the cell wall, do not in
fact stem from the same mechanism.

\begin{table}[h]
\caption{Influences on pinning and energy loss.}
\label{t:exptsum}
\begin{tabular}{|l|l|l|}
\hline
& Effect on & Effect on \\
External change & energy loss\hspace{.3in} & pinning\\
\hline
\hline
smoother cell walls & decreases & usually less\\
\hline
larger cell diameter & decreases & usually more\\
\hline
higher fluid velocity & none & less\\
\hline
more wire vibration & decreases & decreases \\
\hspace*{.1in}(lower T or stronger& &\\
\hspace*{.1in}excitation)& & \\
\hline
\end{tabular}
\end{table}

\section{Mesh}

We suggest a new mechanism for the energy loss during vortex precession, an
interaction with a vortex mesh that covers the cell walls. Such a mesh,
consisting of short vortex lengths pinned on both ends to the cell wall, is
believed to form easily and quickly on container walls, thanks to the
extremely small vortex core size in superfluid helium \cite{Awschalom}. As
the vortex moves along the wall, its end constantly sweeps through the mesh,
reconnecting with mesh vortices as it goes.  We have confirmed the
possibility of energy loss through reconnection in computer simulations
\cite{Ingrid}. Reconnections lead to Kelvin waves along the vortex line.
These waves bring portions of the vortex very close to the wall, leading to
vortex-wall reconnections that shorten the precessing vortex line. At the
low temperatures of our experiment, the Kelvin waves experience so little
damping that many vortex-wall reconnections can result from a single reconnection with a mesh vortex.

This mechanism is qualitatively consistent with several of our observations
about the energy loss during precession.  Smoother cell walls support a
less dense vortex mesh, leading to fewer reconnections and less energy
loss.  A larger cell diameter should also reduce the vortex mesh, since
pinned vortices are less stable near a flatter surface \cite{lukesim}.
The mesh mechanism is also compatible with the similar dissipation
rates observed for precession between $N=2$ and $N=1$ and for precession
between $N=1$ and $N=0$. Although the additional trapped
circulation increases the horizontal speed of the moving vortex, our simulations suggest that the distance along the
wall traveled by the end of the vortex is determined mainly by vertical
oscillations rather than horizontal motion \cite{Ingrid}.  Hence the total
distance traveled, and therefore the rate of encountering mesh vortices,
has little dependence on the fluid velocity in the cell.

\begin{figure}[tbh]
\scalebox{.6}{\includegraphics{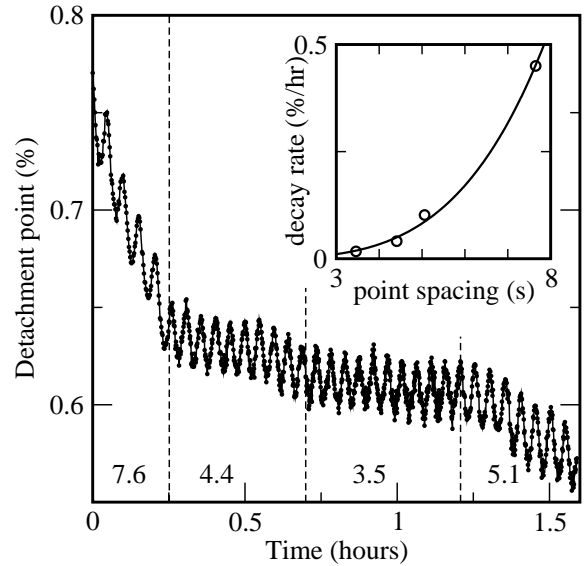}}
\caption{Influence of wire excitation on vortex energy loss.  The intervals
between successive excitations range from 3.5 to 7.6 seconds.  Energy loss
decreases with more frequent excitation, but as shown in the inset it remains
positive; strong excitation of the wire does not increase the energy stored in
the vortex.}
\label{f:hitfast}
\end{figure}

A final point is how the wire's vibration interacts with the precessing
vortex.  As shown in \cite{helpin}, the wire can impart energy to the
vortex line, and increasing the average wire velocity reduces the
observed energy loss from the vortex.  The average wire velocity can be
increased by increasing the initial vibration amplitude, by reducing the
time between excitations, or by cooling the helium to increase the time
constant of the vibration's decay.  All methods have the same effect on the moving
vortex.  One intriguing observation is that the wire excitation never
causes the trapped vortex to gain energy, even though the energy lost
from the wire in each pulse is much larger than that stored in the partially trapped
circulation.  As shown in Figure \ref{f:hitfast},
the dissipation rate decreases as the wire motion increases,
but it appears to level out at zero.
This is consistent with a mesh mechanism of dissipation.  Perturbing the
vortex more strongly must ultimately lead to a steady-state situation
where the energy imparted to the vortex equals the energy removed through
reconnections.  Stronger perturbations of the vortex lead to a longer
steady-state length and increased rate of reconnections.

In our scenario, the pinning comes about for completely different reasons
from the energy loss.  The energy loss depends on the interaction with a
mesh of wall vortices, but there is no obvious reason that the wall
vortices should dominate pinning.  A more likely influence is the
superfluid velocity near a potential pin site, which can sweep a vortex
away and prevent pinning.  Indeed, we see this experimentally for
precession between $N=2$ and $N=1$.  As noted above, the fluid velocity near
the moving vortex is roughly three times as large as in the more common
situation where the vortex precession is between $N=1$ and $N=0$. 
Significantly, we have {\em never} observed pinning at $N>1$, despite more
than seven hours of precession in various wires and more than 35 additional
hours with $N>1$.  Pinning also increases in our larger-diameter cells,
where the $1/r$ falloff of the field from the trapped vortex yields a much
smaller fluid velocity near the cell wall.  By contrast, the energy loss
depends little on the amount of trapped circulation and decreases with
larger cell diameter.

Wall roughness may influence pinning both by providing more
irregularities that can serve as pin sites and also by distorting the
local velocity field.  For rough walls, the fluid has a smaller laminar
flow region, and the resulting increase in local flow velocity could
reduce vortex pinning.  If the stability of a vortex at a pin
site depends both on the strength of the pin (e.g., the size and shape
of a bump on the wall) and on the details of the nearby fluid flow that
might dislodge the vortex, then the dependence of pinning on roughness
could be irregular, as observed.  Rougher walls would increase both
the fluid speed and the typical bump size, but the balance between them
might not have any simple behavior.

Our final probe is the wire vibration itself.  We
reliably observe an increase in pinning with increased temperature.
Pinning also seems to increase when we decrease the vibration amplitude
for the measurements, although our experiments with changing
the amplitude are not extensive.  We conclude that wire vibration
can dislodge vortices from pin sites.

\section{Conclusion}

We measure energy loss and pinning for a vortex extending from a vibrating
wire in the middle of a cylindrical container to the edge of the
container.  Both effects stem from interaction between the vortex and the
surface, but their different dependence on external parameters,
particularly fluid velocity and cell diameter, shows that different
mechanisms must be responsible for the two. We propose that energy loss
arises from reconnections with a mesh of vortices pinned to the cell wall,
which induce further reconnections with the cell wall itself and
deposit segments of the moving vortex on the wall.  Pinning is instead
governed by competition between the degree of roughness on the cell wall
and the fluid velocity field near the wall.

\section{Acknowledgements}
We thank I. Neumann for useful discussions and NSF DMR 0243904 for
funding.

\end{document}